\documentclass[aps,prl,twocolumn,showpacs]{revtex4}

\usepackage{bm}%
\usepackage[colorlinks=true,linkcolor=red]{hyperref}
\usepackage{amsmath}
\usepackage{amsfonts}
\usepackage{amssymb}
\usepackage{verbatim}
\usepackage{epsfig}
\usepackage{graphicx}
\usepackage[sort&compress]{natbib}
\usepackage{times}

\newcommand{\newc}{\newcommand}
\newc{\beq}{\begin{equation}}
\newc{\eeq}{\end{equation}}
\newc{\kt}{\rangle}   
\newc{\br}{\langle}
\newc{\ld}{\lambda}

\begin{document}

\title{\bf\noindent Phase Transitions in the Distribution of Bipartite 
Entanglement of a Random Pure State}
\author{Celine Nadal$^1$, Satya N. Majumdar$^1$ and Massimo Vergassola$^2$}
\affiliation{ $^1$ Laboratoire de Physique Th\'{e}orique et Mod\`{e}les
Statistiques (UMR 8626 du CNRS), Universit\'{e} Paris-Sud,
B\^{a}timent 100, 91405 Orsay Cedex, France. \\
$^2$ Institut Pasteur, CNRS URA 2171, F-75724 Paris 15, France.}

\date{\today}

\begin{abstract}
  Using a Coulomb gas method, we compute analytically the probability
  distribution of the Renyi entropies (a standard measure of
  entanglement) for a random pure state of a large bipartite quantum
  system. We show that, for any order $q>1$ of the Renyi entropy,
  there are two critical values at which the entropy's probability
  distribution changes shape. These critical points correspond to two
  different transitions in the corresponding charge density of the
  Coulomb gas\,: the disappearance of an integrable singularity at the
  origin and the detachement of a single-charge drop from the
  continuum sea of all the other charges. These transitions
  respectively control the left and right tails of the entropy's
  probability distribution, as verified also by Monte Carlo numerical
  simulations of the Coulomb gas equilibrium dynamics.
\end{abstract}


\maketitle

Entanglement is a crucial resource in quantum information and
computation~\cite{NC} as a measure of nonclassical correlations
between different parts of a quantum system. To exploit such
correlations to the maximum advantage in quantum algorithms, it is
desirable to create states with large entanglement.  A potential
candidate for such a state is a bipartite {\it random} pure state, its
{\em average} entanglement entropy being almost
maximal~\cite{Lubkin,Page}.  Such a random pure state can also be used
as a null model or reference point to which the entanglement of an
arbitrary time-evolving quantum state may be compared. In addition,
such random states are also relevant in quantum chaotic or
non-integrable systems~\cite{chaos,BL}.

The conclusion that a random pure state in a bipartite system has near
maximal entropy is based only on the result for the {\em average}
entropy~\cite{Lubkin,Page}.  Even though the average entropy of the
random state may be {\em close} to its maximal value, the probability
of the {\em closeness} may actually be very small (see below).  The
quantitative evaluation of this probability requires to compute the
full probability distribution of the entropy, namely its large
deviation tails.  In addition, the distribution of bipartite
entanglement may also be used to characterize entanglement in a
multipartite system~\cite{Facchi1,Facchi2}. In this Letter we compute
the full distribution of the Renyi entanglement entropies (defined
later) for random pure states of a bipartite system. The calculation
is realized using a Coulomb gas method and is valid in the limit when
both subsystems are large. A by-product of our results is the
behavior of the probability that the entropy approaches its maximal 
value $\ln N$.

We start with a standard bipartite system $A\otimes B$ composed of two
smaller subsystems $A$ and $B$, whose respective Hilbert spaces ${\cal
  H}^{(N)}_A$ and ${\cal H}^{(M)}_B$ have dimensions $N$ and $M$. For
simplicity, we focus on the $N=M$ case, though all our results can be
extended to the $N\ne M$ case. Let $|\psi \kt$ be a normalized pure
state of the full system with its density matrix $\rho = |\psi\kt\,
\br\psi|$ satisfying ${\rm Tr}[\rho]=1$.  The two reduced density
matrices are denoted $\rho_A = {\rm Tr}_B[\rho]$ and $\rho_B = {\rm
  Tr}_A[\rho]$. It is not difficult to prove that both $\rho_A$ and
$\rho_B$ share the same set of non-negative eigenvalues
$\{\lambda_1,\lambda_2,\ldots,\lambda_N\}$ with $\sum_{i=1}^N
\lambda_i=1$. Let $|\lambda_i^{A}\kt$ and $|\lambda_i^{B}\kt$ denote
the respective eigenvectors of $\rho_A$ and $\rho_B$. In this
so-called Schmidt basis, an arbitrary pure state can be represented as
\beq |\psi\kt = \sum_{i=1}^{N} \sqrt{\lambda_i}\, |\lambda_i^A\kt
\otimes |\lambda_i^B \kt.
\label{sch1}
\eeq This representation is very useful for characterizing the
entanglement between $A$ and $B$. For example, consider two opposite
limiting situations: (i) One of the eigenvalues, say $\lambda_i$, is
unity and the remaining $N-1$ are identically zero. Then, $|\psi\kt=
\sqrt{\lambda_i}\, |\lambda_i^A\kt \otimes |\lambda_i^B\kt$ factorizes
and the system is completely ${\it unentangled}$. (ii) All eigenvalues
are equal ($\lambda_i=1/N$ for all $i$). Then, all the states are
equally present in Eq. (\ref{sch1}) and the state $|\psi\kt$ is {\it
  maximally} entangled. A standard measure of entanglement is the von
Neumann entropy, $S_{VN}= -\sum_i \lambda_i \ln \lambda_i$, which takes
its minimum value $0$ in situation (i) and its maximal value $\ln N$
in situation (ii). Another useful measure of entanglement is provided
by Renyi's entropies, the quantities of major interest here\,:
\begin{equation}
\label{Renyi}
S_q=\frac{1}{1-q}\, \ln\left[\sum_{i=1}^N \lambda_i^q\right]\,,
\end{equation}
which also attain their minimum value $0$ in situation (i) and their
maximum value $\ln N$ in (ii). As $q\to 1$ and
$q\to\infty$, the Renyi entropy tends respectively to the von Neumann entropy
$S_{VN}$ and $-\ln \lambda_{\rm max}$, where $\lambda_{\rm max}$ is
the largest eigenvalue.

A pure state $|\psi\kt$ is called {\it random} when it is sampled
according to the uniform Haar measure (the unique unitarily invariant
measure) over the full Hilbert space. As a result, the eigenvalues
$\{\lambda_i\}$'s also become random variables with the joint
distribution (for $M=N$) ~\cite{Page}\,: \beq
P[\{\ld_i\}]\!=\!\frac{1}{Z_0} \prod_{i=1}^{N}
\ld_i^{\frac{\beta}{2}-1}\! \prod_{j<k} |\ld_j-\ld_k|^\beta
\delta\left(\sum_{i=1}^N \ld_i -1 \right)\,.
\label{jpdf1}
\eeq 
Here, $\beta=2$ and the $\delta$-function enforces the unit trace
constraint ${\rm Tr}[\rho_A]=1$.  Apart from this constraint,
(\ref{jpdf1}) is identical to the eigenvalue distribution of random
Gaussian Wishart (covariance) matrices. For random 
matrices, the Dyson index
$\beta$ takes the values $1$, $2$ or $4$ depending on whether the
matrix is real, complex or quaternion. Hence, we shall study
(\ref{jpdf1}) for general $\beta$, even though $\beta=2$ in the
quantum context. The normalization constant $Z_0$ can be computed
exactly using Selberg's integrals~\cite{ZS} as $Z_0\sim e^{-\beta
  N^2/4}$, to leading order in $N$.

Since the $\lambda_i$'s are random variables distributed as in
(\ref{jpdf1}), the von Neumann and the Renyi entropies in
(\ref{Renyi}) are also random variables.  Statistical properties of
these observables, as well as others such as concurrence, purity,
minimum eigenvalue etc., have been studied
extensively~\cite{Lubkin,Page,ZS,Giraud,Znidaric,MBL,Facchi1,Facchi2}.
In particular, the average von Neumann entropy $\langle
S_{VN}\rangle=\ln N -1/2$, is close for large $N$ to its maximum value
\cite{Page}. In contrast, few studies have addressed the full
distribution of the entropy, an exception being the purity
$\Sigma_2=\sum_{i=1}^N \lambda_i^2$\,: for small $N$, the distribution
of purity is known exactly~\cite{Giraud}; for large $N$, the Laplace
transform of the purity distribution was studied recently for positive
Laplace variables~\cite{Facchi2}. However, the inverse Laplace
transform of this quantity provides only partial information about the
purity distribution.

The goal of our Letter is to compute analytically, for large $N$ and
all $q>1$, the full distribution of the Renyi entropies in
(\ref{Renyi}), or equivalently of $\Sigma_q=\sum_{i=1}^N
\lambda_i^q=\exp[(1-q)S_q]$. The quantities $\Sigma_q$ satisfy the
inequalities $N^{1-q}\le \Sigma_q\le 1$ for $q>1$, with the upper and
lower bounds corresponding to the unentangled (i) and the maximally
entangled (ii) situations.  The distribution of $\Sigma_q$ is written
using (\ref{jpdf1}) as
\begin{equation}   
P(\Sigma_q,N)= \int P[\{\lambda_i\}] \delta\left(\sum_i 
\lambda_i^q-\Sigma_q\right)\, \prod_i d\ld_i\,.  
\label{Sqdist1} 
\end{equation} 
The approach we employed to treat (\ref{Sqdist1}) is a saddle-point
method to identify the configuration of the eigenvalues
$\{\lambda_i\}$'s that dominates for large $N$. Configurations at
large $N$ are characterized by the continuous density
$\rho(\lambda,N)= N^{-1}\sum_i \delta(\lambda-\lambda_i)$ and the main
challenge, accomplished here, is to find the saddle-point density
$\rho_c(\lambda,N)$.

\begin{figure} 
\includegraphics[width=.9\hsize]{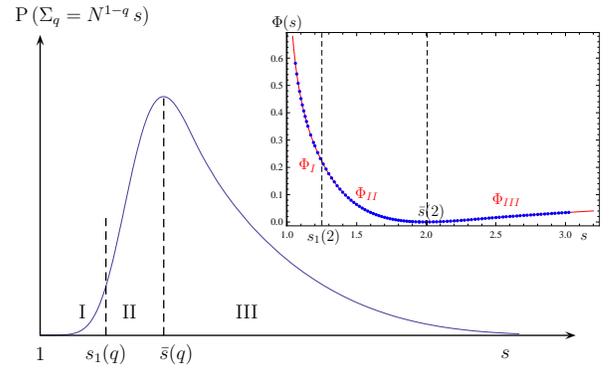} 
\caption{ The schematic distribution of $\Sigma_q= N^{1-q}s
  =\sum_{i=1}^N \lambda_i^q$ as a function of $s$ for fixed large
  $N$. Two critical points $s=s_1(q)$ and $s={\bar s}(q)$ separate
  three regimes $I$, $II$ and $III$ characterized by the different
  optimal densities shown in Fig.~\ref{fig:density}. The maximally
  entangled state $s=1$ is at the extreme left, in the large deviation
  tail well-spaced from the average ${\bar s}(q)$. (Inset) The large
  deviation functions $\Phi$ for the distribution of $\Sigma_q$, in
  the three different regimes. Analytical predictions (red solid line) are
  compared to the results (blue points) of Monte Carlo numerical simulations
  of the Coulomb gas equilibrium dynamics. }
\label{fig:dist1} 
\end{figure}

\begin{figure}  
\includegraphics[width=.9\hsize]{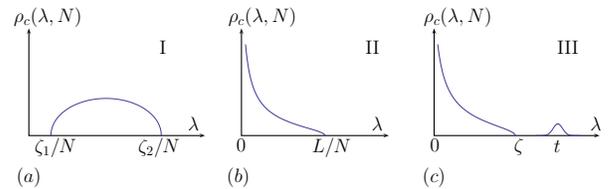}
\caption{ Scheme of the optimal saddle-point density $\rho_c$ of the
  eigenvalues, or equivalently of the Coulomb gas charges, for $1\le
  s\le s_1(q)$ (regime I), $s_1(q)\le s \le {\bar s}(q)$ (II) and $s>
  {\bar s}(q)$ (III).  In the regime III, the maximum eigenvalue
  becomes larger than all the others ($\sim O(1/N)$), as shown by
  the isolated bump at $t$ in (2c).}
\label{fig:density}
\end{figure}

Let us first summarize our main results. The normalization
$\sum_{i=1}^N \lambda_i=1$ implies that the typical amplitude of the
eigenvalues $\lambda_i\sim 1/N$ and hence $\Sigma_q\sim N^{1-q}$ for
large $N$. We define then the rescaled intensive variable $s\equiv
N^{q-1}\,\Sigma_q$ whose lowest value $s=1$ corresponds to the
maximally entangled situation (ii).  In Fig. 1, a typical plot of
$P(\Sigma_q=N^{1-q}s, N)$ {\rm vs} $s$ is shown for large $N$ and
fixed $q>1$\,: the distribution has a Gaussian peak flanked on both
sides by non-Gaussian tails. Specifically, we find two critical values
$s=s_1(q)$ and $s={\bar s}(q)$ separating three regimes I ($1\le s\le
s_1(q)$), II ($s_1(q)\le s\le {\bar s}(q)$) and III ($s> {\bar
  s}(q)$). At the first critical point $s_1(q)$ the distribution has a
weak singularity (third derivative is discontinuous). At the second
critical point ${\bar s}(q)$, a Bose-Einstein type condensation
transition occurs (see below). These changes are a direct consequence
of two phase transitions in the associated optimal charge density
(shown in Fig. 2). In regime I, the optimal charge density has a
compact support $[\zeta_1/N, \zeta_2/N]$, where $\zeta_1$ is strictly
positive (Fig. 2a). When the control parameter $s$ exceeds $s_1(q)$
(regime II), the left edge of the support sticks to zero while the
upper edge $L/N$ moves to the right as $s$ increases (Fig. 2b), till
the second critical value ${\bar s}(q)$ corresponding to $L=4$. For
$s>{\bar s}(q)$ (regime III), we find that one eigenvalue (the small
bump in Fig. 2c) splits off the sea
of all the other $N-1$ eigenvalues, which remain $\sim O(1/N)$.  This second
phase transition is reminiscent of the real-space condensation
phenomenon observed in a class of lattice models for mass transport,
where a single lattice site carries a thermodynamically large
mass~\cite{MEZ}.

Note that for $q=2$, the presence of two phase transitions was also
noticed in Ref.~\cite{Facchi2} for the Laplace transform of the
distribution $P(\Sigma_2,N)$. However, the nature of the regime III
was not elucidated and the corresponding optimal density and the
partition function were not calculated. To derive the full distribution
$P(\Sigma_2, N)$, one needs the partition function in all three
regimes, which is what we do here for all $q>1$ at large $N$. We also
find exact expressions for the two critical points: $4^{-q}{\bar s}(q)=
{\Gamma(q+1/2)}/{\left(\sqrt{\pi}\Gamma(q+2)\right)}$ and $(4(q+1)/3q)^{-q}s_1(q)=
{\Gamma(q+3/2)}/{\left(\sqrt{\pi}\Gamma(q+2)\right)}$.
From the Gaussian form near the peak $s={\bar s}(q)$, we
also read off the mean and the variance of the entropy $S_q$ for all
$q$\,: \begin{equation} \langle S_q\rangle \approx \ln (N)-\frac{\ln
    {\bar s}(q)}{q-1};\quad {\rm Var}(S_q)\approx \frac{q}{2\beta
    N^2}.  \label{meanvar} \end{equation}

Let us now briefly outline how the previous results are derived.  By
using \eqref{jpdf1} and \eqref{Sqdist1}, we obtain
\begin{equation}
P(\Sigma_q,N)=\frac{Z(\Sigma_q)}{Z_0}\,;\; 
Z(\Sigma_q)=\!\int e^{-\beta E(\{\lambda_i\})}\,
\prod_i d\ld_i\,,
\label{pf1}
\end{equation}
with $E(\{\lambda_i\})= -(1/2-1/\beta)\sum_i
\ln(\lambda_i)-\sum_{i<j}\ln |\lambda_i-\lambda_j|$ and the integral
runs over the subspace satisfying the two constraints, $\sum_i
\lambda_i=1$ and $\sum_i \lambda_i^q=\Sigma_q$. The expression for
$E(\{\lambda_i\})$ is interpreted as the energy of a Coulomb gas of
charged particles with coordinates $\lambda_i$ that repel each other
via $2$-d logarithmic interactions and are also subject to an external
logarithmic potential.  In the large $N$ limit, we can characterize
the configuration of the Coulomb gas' charges by the normalized
density $\rho(\lambda,N)= N^{-1}\sum_i \delta(\lambda-\lambda_i)$. Due
to the constraint $\sum_i \lambda_i=1$, typically $\lambda_i\sim
1/N$. Hence, the charge density scales as $\rho(\lambda,N)\approx
N\,\rho(\lambda N)$ and we introduce the rescaled variable $s\equiv
N^{q-1}\Sigma_q$. We then replace the multiple integral in
(\ref{Sqdist1}) by a functional integral over all possible normalized
and rescaled charge density functions $\rho(x)$ satisfying the three
constraints: $\int \rho(x)dx=1$, $\int x\rho(x)dx=1$ and $\int x^q
\rho(x) dx = s$. The resulting functional integral over $\rho(x)$ is
evaluated in the large $N$ limit via the saddle point method.
This constrained Coulomb gas approach has been used successfully in a
variety of contexts that include the distribution of the top
eigenvalues of Gaussian and Wishart matrices~\cite{DM,vivo1,MV}, phase
transition in the restricted trace ensemble~\cite{Akemann}, purity
partition function in bipartite systems~\cite{Facchi2},
nonintersecting Brownian interfaces~\cite{nadal1}, quantum transport
in chaotic cavities~\cite{vivo2}, information and communication
systems~\cite{kaz}, and the index distribution for Gaussian random
fields~\cite{BD,FW} and Gaussian matrices~\cite{nadal2}.

The constrained Coulomb gas approach yields $P\left(\Sigma_q
  =N^{1-q}\: s \right)\propto \int \mathcal{D}\left[ \rho \right] \,
e^{-\beta \, N^2 \, E_s\left[ \rho \right]}$. To the leading order in
$N$, the effective energy reads
\begin{widetext}
\begin{eqnarray}
\label{action}
E_s[\rho]\!=\! 
-\frac{1}{2}\!\int_{0}^\infty\!\!\!\!\int_{0}^\infty\!\!dx\,
dx^\prime \rho(x) \rho(x^\prime)\ln|x-x^\prime|+\!
\mu_0\!\left(\!\int_{0}^\infty\!\! dx\rho(x)-1\!\right)+\! 
\mu_1\!\left(\!\int_{0}^\infty\!\!dx x \rho(x)-1\!\right)+\!
\mu_2\!\left(\!\int_{0}^\infty\!\!dx x^q \rho(x)-s\!\right),
\end{eqnarray}
\end{widetext} 
where the Lagrange multipliers $\mu_0$, $\mu_1$ and $\mu_2$ enforce
the constraints.  For large $N$, the method of steepest descent gives:
$P\left(\Sigma_q=N^{1-q}\: s \right) \propto e^{-\beta N^2 \,
  E_s\left[ \rho_c \right]}$ where $\rho_c(x)$ minimizes the energy\,:
$\delta E_s[\rho]/\delta\rho=0$.  This gives the integral equation
\begin{equation}
  V(x)=\mu_0+ \mu_1 x+ \mu_2 x^q =
 \int_0^{\infty} \rho_c(x^\prime) \ln|x-x^\prime| d x^\prime\,,
\label{saddle}
\end{equation}
with $V(x)$ acting like an effective external potential.
Differentiating once more with respect to $x$ leads to
\begin{equation}
\label{hilbert}
\mu_1 +q \mu_2 x^{q-1}
=\mathcal{P} \int_0^{\infty} \frac{\rho_c(x^\prime)}{x-x^\prime}
dx^\prime\,,
\end{equation}
where $\mathcal{P}$ denotes Cauchy's principal part. The
single-support solution to (\ref{hilbert}) is found by using Tricomi
formula~\cite{Tricomi} and yields the regimes sketched in
Figs.~1 and 2.

{\em Regime I:} For $1\le s\le s_1(q)$, we find that $\mu_1<0$,
$\mu_2>0$ and the effective potential $V(x)$ has a minimum at a
nonzero $x$. This indicates that the charges concentrate around this
nonzero minimum over a support $[\zeta_1,\zeta_2]$ for all $q>1$ (see
Fig.~2a). For $q=2$, the edges $\zeta_{1,2}= 1\mp 2\sqrt{s-1}$ and the
solution $\rho_c(x)= \sqrt{(\zeta_2-x)(x-\zeta_1)}/{\left(2\pi (s-1)\right)}$
vanishes at both edges. This solution exists for $\zeta_1>0$, i.e.,
for $s<s_1(2)= 5/4$, and the distribution $P(\Sigma_2=s/N,N)\sim
e^{-\beta N^2 \Phi_I(s)}$ with the large deviation function
\begin{equation}
\Phi_I(s)= -\frac{1}{4}\ln(s-1) -\frac{1}{8}.   
\label{phi1}
\end{equation}
The behavior for $q\neq 2$ is qualitatively similar, though
the expressions are cumbersome~\cite{details}.  Setting $s=1+\epsilon$
around the maximal entropy state $s=1$, the probability at this
extreme left tail scales as $\sim \epsilon^{\beta N^2/4}$, i.e. it is very
small for large $N$. As $s$ approaches $s_1(q)$ from
below, $\mu_1$ and the minimum of $V(x)$ tend to zero. This signals that
the charges now concentrate near the origin and the onset of
regime II.

{\em Regime II:} For $s_1(q)\le s\le {\bar s}(q)$, the charges
concentrate over a support $[0,L]$ (see Fig. 2b). For $q=2$, the
optimal charge density takes the simple form $\rho_c(x) =
\sqrt{(L-x)}(A+Bx)/{\left(\pi \sqrt{x}\right)}$, where $A=4(L-2)/L^2$, $B=
8(4-L)/L^3$ and the right edge $L=6-2\sqrt{9-4s}$. Evaluating the
energy for large $N$, we get $P(\Sigma_2=s/N,N)\sim e^{-\beta N^2
  \Phi_{II}(s)}$ with
\begin{equation}
\Phi_{II}(s)= -\frac{1}{2}\ln(L/4)+\frac{6}{L^2}-\frac{5}{L}+\frac{7}{8}\,.
\label{phi2}
\end{equation}
Comparing (\ref{phi1}) and \eqref{phi2}, it is verified that the large
deviation functions match at the critical point $s_1=5/4$ up to the
second derivative, while the third is discontinuous:
$\Phi_{I}^{(3)}(5/4)=-32$ and $\Phi_{II}^{(3)}(5/4)=-16$.  The
function $\Phi_{II}(s)$ is quadratic $\Phi_{II}(s)\approx (s-2)^2/8$
around its minimum at $s={\bar s}(2)=2$. Thus, the distribution
$P(\Sigma_2=s/N,N)$ has a Gaussian peak near $s=2$, with the mean
$\langle \Sigma_2\rangle =2/N$ and the variance ${\rm
  Var}(\Sigma_2)=4/(\beta N^4)$ for large $N$. The corresponding
expressions for arbitrary $q>1$ are given in \eqref{meanvar}.

For any $q>1$, $\mu_1$ is positive, $\mu_2\to 0$ as $s\to {\bar s}(q)$
and $\mu_2<0$ for $s>{\bar s}(q)$. This indicates that the
potential $V(x)$ in \eqref{saddle} becomes non-monotonic for $s>{\bar
  s}(q)$\,: it increases around the origin, reaches a maximum at $x^*=
(-\mu_1/{q\mu_2})^{1/(q-1)}$ and then decreases monotonically for
$x>x^*$. It follows that the minimum at $x=0$ as well as the
associated saddle-point solution become {\em metastable}. This
solution still exists over a finite range for $s>{\bar s}(q)$
(e.g., for $q=2$ over $2\le s\le 9/4$). For $s>{\bar s}(q)$,
however, there is a second {\em stable} solution where one eigenvalue
splits off the sea of the remaining $(N-1)$ eigenvalues (see regime
III below).  For $s>{\bar s}(q)$, the energy associated with this
stable solution is lower by $\sim O(1/N)$ only, as compared to the
energy of the metastable solution.

{\em Regime III:} For $s>{\bar s}(q)$, the correct density of states
consists of two disjoint parts: (a) $(N-1)$ eigenvalues remain $\sim
O(1/N)$, concentrated over a finite support including the origin; (b)
the top eigenvalue $\lambda_{\rm max}$ takes a larger value and moves
away from the sea of all the others (see Fig.~2c).  The saddle point
method thus needs to be slightly revised.  For simplicity, let us
focus only on $q=2$. We write $\lambda_{\rm max}=t$ and label the
remaining $(N-1)$ eigenvalues by their continuous density
$\rho(\lambda)=\frac{1}{N-1}\sum_{i \neq max}
\delta(\lambda-\lambda_i)$. We then express the energy
$E[\{\lambda_i\}]$ in \eqref{pf1} in terms of $\rho(\lambda)$ and $t$,
treating both of them as variables. This gives $P(\Sigma_2=S,N)\propto
\int \mathcal{D} \rho \int dt \; e^{-\beta H_S[\rho,t]}$, where the
effective energy $H_S[\rho,t]$ has a long expression that includes
$\rho$, $t$ and three Lagrange multipliers enforcing the
constraints~\cite{details}. Assuming that $\rho(\lambda)$ has a finite
support over $[0,\zeta]$ with $\zeta<t$, we minimize the effective
energy over both $\rho$ and $t$. The equations $\delta H_S/{\delta
  \rho}=0=\partial H_S/{\partial t}$ are solved again using Tricomi's
theorem \cite{Tricomi}.  Substituting the solutions for
$\rho(\lambda)$ and $t$ in the effective energy finally yields the
distribution $P(\Sigma_2,N)$ at the leading order in $N$. We have
verified that in the regime $2\le s\le 9/4$, the resulting
distribution coincides with that of regime II, i.e. the transition at
$\Sigma_2\to 2/N$ is smooth.  The maximum eigenvalue $\lambda_{\rm
  max}=t$ dominates at the upper edge of the regime III, when
$\Sigma_2\sim O(1)$, and we find~\cite{details} that
$P(\Sigma_2=S,N)\sim (1-\sqrt{S})^{\beta N^2/2}$.

{\em Numerical Simulations.} To verify analytical predictions, we simulated
the distribution (\ref{jpdf1}) of the eigenvalues $\lambda_i$, which is interpreted as the
Boltzmann weight of a Coulomb gas and sampled using
the Metropolis algorithm (see, e.g., \cite{Krauth}).  Specifically, we
start with a configuration of the $\lambda_i$'s satisfying $\sum_i
\lambda_i=1$. The moves in the Metropolis scheme consist of picking at
random a pair $(\lambda_i, \lambda_j)$ and proposing to modify them as
$(\lambda_i+\varepsilon, \lambda_j-\varepsilon)$ where $\varepsilon$
is set to achieve the standard average rejection rate $1/2$
\cite{Krauth}. As usual, the move is accepted with probability
$e^{-\beta \Delta E}$ if $\Delta E>0$ and with probability $1$ if
$\Delta E<0$, where $\Delta E$ is the change in energy
$E[\{\lambda_i\}]$ (the move is rejected if one of the eigenvalues
becomes negative). This ensures that at long times we reach thermal
equilibrium with the correct Boltzmann weight $\propto e^{-\beta
  E[\{\lambda_i\}]}$ satisfying the constraint $\sum_i \lambda_i=1$.  We then construct the histogram of $P(\Sigma_q=\sum_i \lambda_i^q,N)$. Numerical data compare very well with our analytical predictions (see the inset of Fig.~1 for $q=2$, $N=50$)\,; we also verified that a single eigenvalue detaches from the sea in regime III (intuitively, multiple drops are unfavorable as they compress the sea more than a single drop due to the convexity of $\sum\lambda_i^q$ for $q>1$).

In conclusion, we have obtained the first complete characterization of
the quantum entanglement's statistical properties in a bipartite
random pure state of large dimensions $N$.  The average of the Renyi
entropies is indeed close to its largest value $\ln N$. This is,
however, the mere consequence of the typical amplitude $1/N$ of the
density matrix eigenvalues. The distribution of the
eigenvalues mostly affects the $O(1)$ contribution to the entropy.
The probability to approach $\ln N$ is actually found to decay
rapidly at large $N$, as clearly shown by the full probability
distribution of the Renyi entropies derived here. The spreading of the
eigenvalues becomes prominent in the regime III (in Figs.~1 and 2)
where a condensation occurs and the contribution by the single top
eigenvalue of the density matrix is thermodynamically relevant.

{\bf Acknowledgements} We thank O. Bohigas and A. Scardicchio for
useful discussions.


\begin{thebibliography}{99}

\bibitem{NC} M.A. Nielsen and I.L. Chuang, {\it Quantum Computation
and Quantum Information} (Cambridge Univ. Press, Cambridge, 2000).

\bibitem{Lubkin} E. Lubkin, J. Math. Phys. {\bf 19}, 1028 (1978); S. Lloyd
and H. Pagels, Ann. Phys. (N.Y.) {\bf 188}, 186 (1988).

\bibitem{Page} D.N. Page, Phys. Rev. Lett. {\bf 71}, 1291 (1993).

\bibitem{chaos} O. Bohigas, M. J. Giannoni and C. Schmit,
Phys. Rev. Lett. {\bf 52}, 1 (1984).

\bibitem{BL} J.N. Bandyopadhyay and A. Lakshminarayan, Phys. Rev. Lett. {\bf 89}, 
060402 (2002) and references therein.

\bibitem{Facchi1} P. Facchi, G. Florio, G. Parisi and S. Pascazio, Phys. Rev. A
{\bf 77}, 060304 (R) (2008).

\bibitem{Facchi2} P. Facchi et al., Phys. Rev. Lett. {\bf 101}, 050502 (2008).

\bibitem{ZS} K. Zyczkowski and H-J. Sommers, J.
Phys. A: Math. Gen. {\bf 34}, 7111 (2001).

\bibitem{Giraud} O. Giraud, J. Phys. A.: Math. Theor. {\bf 40}, 1053 (2007).

\bibitem{Znidaric} M. Znidaric, J. Phys. A: Math. Theor. {\bf 40}, F105 (2007).
 
\bibitem{MBL} S.N. Majumdar, O. Bohigas, 
and A. Lakshminarayan, J. Stat. Phys. {\bf  
131}, 33 (2008).

\bibitem{MEZ} S.N. Majumdar, M.R. Evans and R.K.P. Zia, 
Phys. Rev. Lett. {\bf 94}, 
180601 (2005).

\bibitem{DM} D.S. Dean and S.N. Majumdar, Phys. Rev. Lett. {\bf 97}, 160201
(2006); Phys. Rev. E {\bf 77}, 41108 (2008).

\bibitem{vivo1} P. Vivo, S.N. Majumdar and O. Bohigas,
J. Phys. A: Math. Theor. {\bf 40}, 4317 (2007).

\bibitem{MV} S.N. Majumdar and M. Vergassola, 
Phys. Rev. Lett. {\bf 102}, 060601 (2009).

\bibitem{Akemann} G. Akemann et al., Phys. Rev. E {\bf 59}, 1489 (1999).

\bibitem{nadal1} C. Nadal and S.N. Majumdar, Phys. Rev. E {\bf 79}, 
061117 (2009).

\bibitem{vivo2} P. Vivo, S.N. Majumdar and 
O. Bohigas, Phys. Rev. Lett. {\bf 101}, 216809 (2008); arXiv:0909:2974 
(2009).

\bibitem{kaz} P. Kazakopoulos et al., [arXiv:0907.5024] (2009).

\bibitem{BD} A.J. Bray and D.S. Dean, Phys. Rev. Lett. {\bf 98}, 150201 
(2007).

\bibitem{FW} Y.V. Fyodorov and I. Williams, J. Stat. Phys. {\bf 129}, 1081 
(2007).

\bibitem{nadal2} S.N. Majumdar et al., [arXiv:0910:0775] (2009).

\bibitem{Tricomi} F.G. Tricomi, {\it Integral Equations} (Pure Appl. Math. V, 
Interscience, London, 1957).

\bibitem{details} Details will be published elsewhere.

\bibitem{Krauth}
W. Krauth, {\it Statistical Mechanics: Algorithms and Computation} (Oxford Univ. Press, Oxford, 2006). 

\end{thebibliography}
\end{document}